\documentclass[twocolumn]{emulateapj}
\usepackage{color}
\usepackage{bm}
\usepackage{amsmath,amssymb}
\usepackage{lineno}

\shorttitle{Modeling of Gamma-ray Flares in PKS\,1510--089}
\shortauthors{Saito et al.}

\usepackage{version}	
\begin{document}
\title{Time-Dependent Modeling of Gamma-ray Flares in Blazar PKS\,1510--089}
\author{S. Saito\altaffilmark{1,\,2}, \L . Stawarz\altaffilmark{3,\,2},
Y. T. Tanaka\altaffilmark{4}, T. Takahashi\altaffilmark{2},
M. Sikora\altaffilmark{5}, R. Moderski\altaffilmark{5}}
\email{\texttt{s.saito@rikkyo.ac.jp}}
\altaffiltext{1}{Department of Physics, Rikkyo University, 3-34-1
Nishi-Ikebukuro, Toshima-ku, Tokyo 171-8501, Japan}
\altaffiltext{2}{Institute of Space and Astronautical Science JAXA, 3-1-1 Yoshinodai, Chuo-ku, Sagamihara, Kanagawa 252-5210, Japan}
\altaffiltext{3}{Astronomical Observatory, Jagiellonian University, ul. Orla 171, 30-244 Krak\'ow, Poland}
\altaffiltext{4}{Hiroshima Astrophysical Science Center, Hiroshima University, 1-3-1 Kagamiyama, Higashi-Hiroshima 739-8526, Japan}
\altaffiltext{5}{Nicolaus Copernicus Astronomical Center, Bartycka 18,
00-716 Warsaw, Poland}

\begin{abstract}
 Here we present a new approach for constraining luminous blazars,
 incorporating fully time-dependent and self-consistent modeling of
 bright $\gamma$-ray flares of PKS\,1510-089 resolved with
 \textit{Fermi}-LAT, in the framework of the internal shock
 scenario. The results of our modeling imply the location of the
 $\gamma$-ray flaring zone outside of the broad-line region, namely
 around $\simeq 0.3$\,pc from the core for a
 free-expanding jet with the opening angle $\Gamma \,
 \theta_\mathrm{jet} \simeq 1$ (where $\Gamma$ is the jet bulk Lorentz
 factor), up to $\simeq 3$\,pc for a collimated outflow with $\Gamma \,
 \theta_\mathrm{jet} \simeq 0.1$. Moreover, under the $\Gamma \,
 \theta_\mathrm{jet} \simeq 1$ condition, our modeling indicates the
 maximum efficiency of the jet production during the flares,
 with the total jet energy flux strongly dominated by protons and
 exceeding the available accretion power in the source. This is in
 contrast to the quiescence states of the blazar, characterized by lower
 jet kinetic power and an approximate energy equipartition between
 different plasma constituents. We demostrate how strictly simultaneous
 observations of flaring PKS\,1510--089 at optical, X-ray, and GeV
 photon energies on hourly timescales, augmented by extensive
 simulations as presented in this paper, may help to impose further
 precise constraints on the magnetization and opening angle of the
 emitting region. Our detailed modeling implies in addition that a
 non-uniformity of the Doppler factor across the jet, caused by the
 radial expansion of the outflow, may lead to a pronounced time
 distortion in the observed $\gamma$-ray light curves, resulting in
 particular in asymmetric flux profiles with substantially extended
 decay phases.

\end{abstract}


\section{Introduction}
\label{S:intro}

Blazars constitute a population of radio-loud Active Galactic Nuclei (AGN) with relativistic jets directed toward the Earth. The observed spectrum of a blazar is dominated by the jet component as a result of a strong relativistic beaming involved, and consists of the two broad humps in the $\nu - \nu F_{\nu}$ representation (hereafter the `spectral energy distribution', SED). The low energy hump, peaked in the infrared--to--X-ray frequency range, is due to the synchrotron emission of ultra-relativistic jet electrons, while the high energy component, peaking in the $\gamma$-ray band, originates from the inverse Compton radiation of relativistic electrons interacting with ambient seed photons produced either externally to the jet (predominantly via a re-procession of the accretion disk emission by the circumnuclear gas and dust), or within the outflow by the synchrotron process.

Blazars can be classified as either BL Lacertae objects (BL Lacs) or flat-spectrum radio quasars (FSRQs), based on the properties of the optical line emission. In particular, the latter class of sources is characterized by the presence of prominent broad and narrow emission lines in their spectra. FSRQs constitute the most luminous population of blazars, with super massive black holes (SMBHs) accreting at high rates \citep[over 1\% of the Eddington limit; e.g.,][]{Sbarrato14}. The available huge accretion mass flux is converted very efficiently to the jet energy flux within the ergospheres of rapidly spinning SMBHs at the expense of the black hole rotational energy \citep{Blandford77,Sikora07,Tchekhovskoy11}. In this context, high-energy observations are indispensable for imposing meaningful constraints on the jet energetics, since the bulk of the radiative energy of luminous blazars is released in the MeV/GeV range \citep[see][]{Ghisellini14}. And indeed, it was repeatedly shown that the amount of the jet power dissipated radiatively during the flares of FSRQs in $\gamma$-rays is often comparable to the corresponding disk luminosities \citep{Tanaka11,Saito13}.
  
Blazars display complex variability patterns on various timescales across the entire electromagnetic spectrum, with high-amplitude flux changes pronounced most clearly at high frequencies. The all-sky $\gamma$-ray survey with the Large Area Telescope onboard the {\it Fermi} satellite (hereafter {\it Fermi}-LAT) unveiled in particular very rapid flares of FSRQs in the GeV range, with the observed flux doubling timescales of the order of hours, down to even sub-hour domain in a few cases \citep[see the analysis of the LAT data for PKS\,1510--089, 3C\,273, and 4C\,21.35 by][]{Foschini11,Foschini13,Saito13,Brown13,Rani13}. Such a dramatic $\gamma$-ray variability indicates a very efficient energization of the jet electrons taking place in compact regions of the enhanced energy dissipation, followed by the rapid radiative cooling of the accelerated particles.

The exact site and structure of those energy dissipation regions is, however, still controversial, in spite of the extensive broad-band observations carried out over the last two decades. This reflects in a wide range of the estimated position of the blazar emission zone along the jet, even for the same object, based on different sets of observables interpreted in the frameworks of different jet emission models. For example, as for PKS\,1510--089, the combined radio, optical polarization, and $\gamma$-ray monitoring suggested the dominant emitting region to be located as far as $\sim 10$\,pc from the central SMBH \citep[the ``far-dissipation zone'' scenario; e.g.,][]{Marscher10,Orienti13}. Meanwhile, another studies based on the SED modeling including the UV, X-ray, and $\gamma$-ray data, placed the blazar zone in the source at sub-pc distances from the central engine \citep[the ``near-dissipation zone'' scenario; e.g.,][]{Kataoka08,Ghisellini10}. Complex multi-zone/multi-component models were also proposed in order to explain broad-band variability and spectral properties of PKS\,1510--089 \citep{Nalewajko12a,Aleksic14}.

In principle, timing characteristics of high-amplitude blazar flares can offer some clues on the location of the active emission zone. First, due to the causality requirement, the linear size of the emission region may be constrained as $r \lesssim c \, \delta \, \tau_d /(1+z)$, for the observed flux \emph{doubling} timescale $\tau_d$ and the Doppler factor of the emitting plasma $\delta$. Assuming further a conical, uniform, and free-expanding jet, whose (small) half-opening angle $\theta_\mathrm{jet} \simeq r/R$ is approximately the inverse of the jet bulk Lorentz factor $\Gamma$, the location of the flaring zone from the central SMBH reads as $R \lesssim c \, \Gamma \, \delta \, \tau_d /(1+z)$. The flux doubling timescale of about a few hours, for example, often inferred for FSRQs from the LAT data as mentioned above, gives therefore $R \lesssim 10^{16}$\,cm for the typically expected in blazar sources $\Gamma \simeq \delta \sim 10$, or $R < 10^3 \, R_S$ in the units of the Schwarzschild radius $R_S$ for the black hole mass of the order of $10^8 \, M_{\odot}$. This would be then consistent with the ``near-dissipation zone'' scenario.

On the other hand, the most recent detections of very high energy $\gamma$-rays (photon energies $>100$\,GeV) from some FSRQs, including PKS\,1510--089 \citep{Abramowski13,Aleksic14,Barnacka2014}, have challenged the above conclusion. That is because such $\gamma$-rays are expected to undergo an efficient annihilation in the interactions with lower-energy circumnuclear photon fields, and in particular with the UV disk emission re-processed within the ``broad line region'' (BLR), if produced at the scales below the characteristic radius of the BLR clouds \citep[$\sim 0.1$\,pc for the typical disk luminosity of the order of $10^{46}$\,erg\,s$^{-1}$; see, e.g.,][]{Moderski2003,Liu06,Tavecchio09,Barnacka2014}. The detection of very high energy $\gamma$-rays from FSRQs seems therefore to require the dominant emission region to be located at larger distances from the core, at least in the framework of the one-zone emission models.

\begin{table}[!t]
{\footnotesize
\noindent
\caption{$\gamma$-ray flaring events of PKS\,1510--089 modeled in this work}
\label{T:flares}
\begin{center}
\begin{tabular}{cccc}
\hline\hline
Name & MJD & $F_{>100\,{\rm MeV}}$ & $\Gamma_{\gamma}$\\
 & (1) & (2) & (3)\\
\hline
Flare\,1 & 55853.5--55854.5 & $14.86 \pm 0.89$ & $1.97 \pm 0.04$\\
Flare\,2 & 55872--55874 & $8.39 \pm 0.44$ & $2.19\pm 0.04$\\                               
\hline\hline
\end{tabular}
\end{center}
(1) Dates of the $\gamma$-ray flux maxima in the daily-binned light curve; (2) photon fluxes measured at the flux maxima in the units of [$10^{-6}$\,ph\,cm$^{-2}$\,s$^{-1}$], averaged over the specified time intervals; (3) the corresponding photon indices. All the values are taken from \citet{Saito13}.}
\end{table}

An important issue, which is currently overlooked in the blazar modeling, but which, in fact, is crucial for constraining luminous blazars, is the exact $\gamma$-ray spectral evolution during the rapid flux changes, along with the exact time profiles of the flares. {\it Fermi}-LAT observations of the brightest FSRQs have revealed that spectral breaks appear around a few/several GeV during the enhanced activity states, and that the low- and high-energy spectral slopes below and above the break, respectively, often change significantly on the timescales of less than a day \citep{Abdo09,Abdo11,Tanaka11,Stern11,Rani13}. This indicates that the $\gamma$-ray data integrated over longer periods of time, which are typically utilized in the blazar SED modeling, should be considered as a superposition of different spectra produced at different stages of the jet evolution, and as such may hardly be used for any exact spectral diagnosis. Of course, in the overwhelming majority of cases long integration LAT exposures times are unavoidable, due to a limited photon statistics. Still, the caution should be kept in mind.

This situation has motivated us to attempt a detailed, \emph{time-dependent}, and fully \emph{self-consistent} modeling of the spectral evolution of bright $\gamma$-ray flares in FSRQs, for which the time profiles can be resolved and the spectra can be constrained on the hourly timescales. In particular, in this paper we present the extensive modeling of the extremely bright $\gamma$-ray flares in PKS\,1510--089, which were analyzed before by \citet[see also \citealt{Brown13} and \citealt{Foschini13}]{Saito13}. This allows us to put uniquely robust (though still model-dependent, to some extent) constraints on the structure and the location of the flaring emission zone in the source, as discussed further below.

Throughout the paper we assume the $\Lambda$CDM cosmology with $\Omega_{\Lambda} = 0.73$, $\Omega_{\rm M} = 0.27$, and $H_{\rm 0} = 71$\,km\,s$^{-1}$\,Mpc$^{-1}$, so that the redshift of PKS\,1510--089, $z=0.361$, corresponds to the luminosity distance of $d_{\rm L} \simeq 1.91$\,Gpc.

\section{Selection of the {\it Fermi}-LAT Data}
\label{S:data}

\begin{table*}[!t]
\begin{center}
\caption{Model parameters of the ``free-expanding jet'' fit to the $\gamma$-ray flares of PKS\,1510--089.}
\label{T:modelparam}
\begin{tabular}{lccc}
\hline\hline
Model parameter & Flare\,1 & Flare\,2 & reference \\
\hline
Minimum electron Lorentz factor, $\gamma_\mathrm{min}$ & \multicolumn{2}{c}{1}  & \citet{Barnacka2014}\\
Break electron Lorentz factor, $\gamma_\mathrm{br}$ & \multicolumn{2}{c}{900} & '' \\
Maximum electron Lorentz factor, $\gamma_\mathrm{max}$ & \multicolumn{2}{c}{$1\times10^5$} & '' \\
Low-energy electron injection index, $p$ & \multicolumn{2}{c}{1.2} & '' \\
High-energy electron injection index, $q$ & \multicolumn{2}{c}{3.4} & '' \\ 
Bulk Lorentz factor of the emitting shell, $\Gamma$ & \multicolumn{2}{c}{22} & '' \\
Jet opening angle, $\theta_\mathrm{jet}$ & \multicolumn{2}{c}{2.6\,deg} & '' \\
Jet viewing angle, $\theta_\mathrm{obs}$ & \multicolumn{2}{c}{2.6\,deg} & '' \\
Jet magnetic field intensity at $10^{18}$\,cm, $B_0$ & \multicolumn{2}{c}{0.75\,G} & '' \\
Characteristic scale of the BLR, $R_{\mathrm{BLR}}$ & \multicolumn{2}{c}{$0.12\times 10^{18}$\,cm} & '' \\
Central energy density of the BLR photon field, $u_{\mathrm{BLR}}$ & \multicolumn{2}{c}{0.06\,erg\,cm$^{-3}$} & '' \\
Characteristic energy of the BLR photons, $h\nu_{\mathrm{BLR}}$ & \multicolumn{2}{c}{10\,eV} & '' \\
Characteristic scale of the HDT, $R_{\mathrm{HDT}}$ & \multicolumn{2}{c}{$1.94\times 10^{18}$\,cm} & '' \\
Central energy density of the HDT photon field, $u_{\mathrm{HDT}}$ & \multicolumn{2}{c}{$5\times 10^{-4}$\,erg\,cm$^{-3}$} & '' \\
Characteristic energy of the HDT photons, $h\nu_{\mathrm{HDT}}$ & \multicolumn{2}{c}{0.15\,eV} & '' \\
Normalization of the electron injection function, $K_e$ & $1.6\times 10^{47}$\,s$^{-1}$ & $0.6\times 10^{47}$\,s$^{-1}$ & this work, \S\,\ref{S:freejet} \\
Distance where the injection starts, $R_\mathrm{start}$ & $0.7\times 10^{18}$\,cm & $2.3\times 10^{18}$\,cm & '' \\
Distance where the injection terminates, $R_\mathrm{stop}$ & $0.9\times 10^{18}$\,cm & $3.4\times 10^{18}$\,cm & '' \\
Distance where the simulation stops, $R_\mathrm{end}$ & $2.3\times 10^{18}$\,cm & $6.9\times 10^{18}$\,cm & '' \\
 \hline
\end{tabular}
\end{center}
\end{table*}

The {\it Fermi}-LAT provides the most complete and sensitive coverage of the $\gamma$-ray sky up to date. Since 2008 it surveys the entire sky every three hours within the photon energy range $20\,\mathrm{MeV}-300\,\mathrm{GeV}$, so that each astrophysical $\gamma$-ray source --- majority of which are blazars \citep{Nolan12} --- is exposed for about 30\,min during one scan. Due to the limited photon statistics, the selection of the {\it Fermi}-LAT data for this work has to focus inevitably on the flaring states of the brightest FSRQs, for which meaningful spectral analysis could be performed with the minimum LAT time resolution. We further restrict our modeling to the well-resolved $\gamma$-ray flares of PKS\,1510--089 which occurred in 2011, as summarized in Table\,\ref{T:flares} following the analysis of \citet{Saito13}. These two flares constitute the best known examples of the prominent, isolated, and coherent events, unlike the majority of the observed blazar flux enhancements which seem rather like a superposition of distinct (though possibly related) but just unresolved sub-flaring/flickering. 

The question remains if the minimum (orbital) 3-hour binning of the LAT data is sufficiently short for resolving the PKS\,1510-089 flares properly. We note in this context that the stochastic modeling of the LAT light curves for the brightest blazars by \citet{Sobolewska14} revealed some hints for sub-hour variability timescales in only four sources, \emph{not} including PKS\,1510--089. These characteristic variability timescales regarded the features of the $\gamma$-ray power spectral density functions, and not the flux doubling timescales studied by \citet{Saito13} or \citet{Brown13}. Taking therefore into account that the rising and decaying phases of the two flares selected for our analysis are very well defined, with no sub-structure appearing even when studied with the shortest binning \citep{Saito13}, we conclude that one can indeed consider a single emitting component to be responsible for the observed flux evolution during the targeted activity epochs of the source, and hence that a meaningful time-dependent modeling can be performed in the framework of a given jet model. The issue of sub-orbital blazar variability in the LAT data will be analyzed in detail in the forthcoming paper Saito et al. (2015, in prep.).

\section{The Model}
\label{S:model}

We simulate the time evolution of the observed source spectrum during the selected flaring events utilizing the \texttt{BLAZAR} code, which was developed by \citet{Moderski2003,Moderski2005} based on the widely anticipated scenario of internal shocks formed by colliding blobs of the jet plasma \citep[e.g.,][see \S\,\ref{S:discussion} for the detailed model description]{Sikora94,Spada01,Boettcher10,Mimica12,Rueda14}. The model can be however applied to a more general situation as well, since it only approximates the flaring emission region with a uniform and expanding shell of the emitting plasma moving along the outflow, and follows self-consistently the evolution of the radiating particles injected into the shell, but does not involve any particular assumption on the exact particle acceleration process involved.

\begin{figure*}[!t]
\begin{center}
 \includegraphics[width=\textwidth,bb=20 10 970 490,clip]{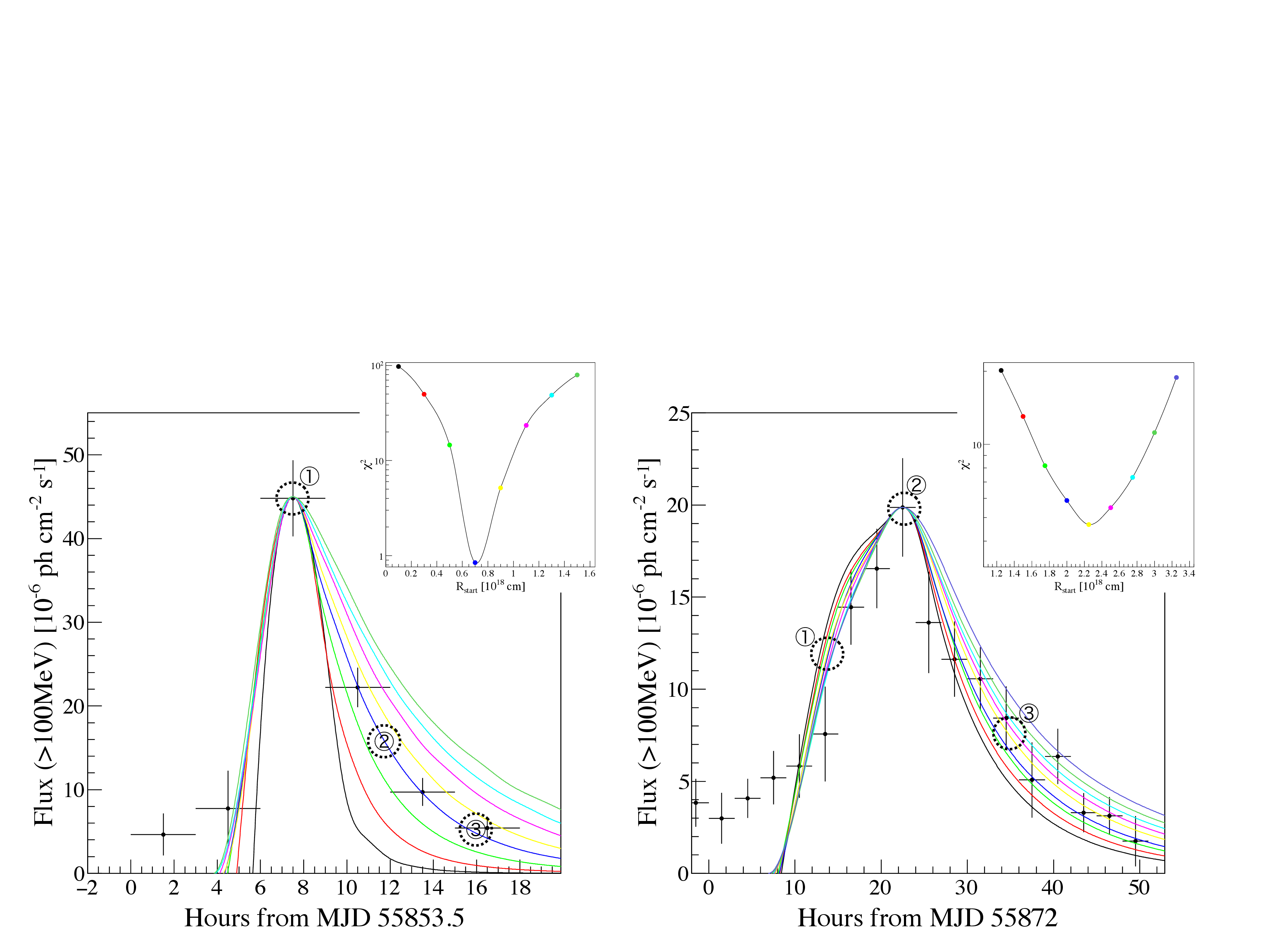}
 \caption[]{Simulated profiles of the analyzed Flare\,1 and Flare\,2 in
 PKS\,1510--089 within the 0.1--300\,GeV range (left and right panels,
 respectively), superposed on the 3-hour binned LAT light curve of the
 source taken from \citet{Saito13}. Simulations denoted in the figure by
 various color curves were performed for different locations of the
 emitting region along the free-expanding outflow, as explained in the
 text (see \S\,\ref{S:freejet}). The $\chi^2$ values for the
 corresponding model fits to the decaying phases of the flares are
 inserted as small figures in the panels. The resulting best-fit
 parameters are given in Table\,\ref{T:modelparam}. The broad-band
 emission spectra presented in Figure\,\ref{F:seds} below coincide with
 the instants marked as ``1'', ``2'', and ``3'' in the panels. We note
 the long-term averaged flux level of the source is
 $2.7\times10^{-7}\,\mathrm{ph\,cm^{-2}\,s^{-1}}$ above 100\,MeV, according to
the {\textit Fermi}-LAT 3FGL catalog \citep{Acero15}.}
\label{F:lcmodelf1}
\end{center}
\end{figure*}

More specifically, in the model the electron injection is assumed to take place at a constant rate with a fixed injection spectrum for a given instance of time when the shell moves from $R_\mathrm{start}$ to $R_\mathrm{stop}$, and to be negligible otherwise. The time evolution of the injected electron energy distribution is calculated in the emitting plasma rest frame (denoted below by primes) following the standard kinetic equation
\begin{equation}
\frac{\partial N_{\gamma}}{\partial t^{\prime}} = -\frac{\partial}{\partial\gamma}\left(N_{\gamma} \, \frac{\mathrm{d}\gamma}{\mathrm{d}t^{\prime}}\right) + Q_{\gamma} \, ,
\label{eq:kin}
\end{equation}
where $N_\gamma$ is the comoving number density of electrons with Lorentz factor $\gamma$, and $Q_{\gamma}$ is the electron injection function. In this paper we do not prime the electron Lorentz factors $\gamma$, noting instead that these always refer to the emitting plasma rest frame. Radiative energy losses due to the synchrotron emission and the inverse-Compton radiation, as well as the adiabatic losses due to the radial expansion of the emitting shell, are all taken into account self-consistently in the $\mathrm{d}\gamma/\mathrm{d}t^{\prime}$ term. For those, the jet magnetic field is assumed to decrease along the outflow as $B^{\prime}\!(R) \propto R^{-1}$, corresponding to the case of the conserved magnetic energy flux with the dominant toroidal component. The energy densities of the external target photon fields for the inverse-Compton scattering, namely of the BLR and of the hot dusty torus (HDT), are assumed to be isotropic in the observer frame and to scale with the distance from the core as
\begin{equation}
u^{\prime}_\mathrm{ext}(R) = \frac{\Gamma^2 \, L_\mathrm{ext}}{4 \pi \, c \, R_\mathrm{ext}^2} \, \frac{1}{1+(R/R_\mathrm{ext})^2} \, ,
\end{equation}
where $R_\mathrm{ext}$ is the characteristic radius and $L_\mathrm{ext}$ represents the total luminosity of a given field \citep[see][]{Sikora09}\footnote{ See in this context \citet{Janiak15}, Appendix A1 therein, for the alternative scaling of the BLR energy density with the distance from the central engine.}. We note that the \texttt{BLAZAR} code offers the correct treatment of the quantum (Klein-Nishina) effects when calculating the inverse-Compton emission \citep[see the discussion in][]{Moderski2005}. 

We parametrize the electron injection spectrum by a broken power-law 
\begin{equation}
Q_{\gamma} = K_e \, \gamma^{-p} \, \left[1+\left({\gamma \over \gamma_\mathrm{br}}\right)^4\right]^{\frac{1}{4} \, (p-q)} \, , 
\end{equation}
where $K_e$ is the normalization parameter, $p$ and $q$ are the low- and high-energy injection indices, respectively, and $\gamma_\mathrm{br}$ is the criticial/break electron Lorentz factor. The injection function is defined within the range of $\gamma_\mathrm{min} \leq \gamma \leq \gamma_\mathrm{max}$.

After simulating the evolution of the electron energy distribution while the shell propagates along the jet based on the equation of balance (\ref{eq:kin}), the observed SED at a given instance of time is calculated by integrating radiative contributions from all the cells located at different radial distances within the shell over the viewing angle $\theta_{\rm obs}$. The corresponding $\gamma$-ray light curve is obtained by extracting the $\gamma$-ray flux at each step of the simulation \citep[see][for the full description]{Moderski2003}.

\section{Time-dependent Modeling}
\label{S:modeling}

Several input parameters for the time-dependent simulations of the analyzed flares were selected based on the recent broad-band fitting of the PKS\,1510--089 spectrum during the 2009 flaring state by \citet{Barnacka2014}. The main characteristics of the external photon fields (BLR and HDT, in particular) were fixed following the detailed studies of the accretion disk emission in the source by \citet{Nalewajko12a}. All of these are summarized in Table\,\ref{T:modelparam}. The jet opening and viewing angles were at first assumed as $\theta_\mathrm{jet} = \theta_\mathrm{obs} = 1/\Gamma$, where $\Gamma$ is the bulk Lorentz factor of a moving shell (\S\,\ref{S:freejet}); in the following steps of the modeling we also considered smaller values of $\theta_\mathrm{jet}$, in accord with the most recent results of the high-resolution radio observations of blazar jets (see \S\,\ref{S:collimatedjet} and references therein).

\subsection{``Free-Expanding Jet'' Model}
\label{S:freejet}

The critical distances along the outflow marking the onset and the termination of the electron injection, $R_\mathrm{start}$ and $R_\mathrm{stop}$, respectively, i.e. the two crucial free parameters of our modeling, were constrained under the condition of a free-expanding outflow $\theta_\mathrm{jet} = \theta_\mathrm{obs} = 1/\Gamma = 2.6$\,deg in the following way. First, the interval $\Delta R = R_\mathrm{stop} - R_\mathrm{start}$ was determined from the observed rising time of the flare $\tau_\mathrm{fl}$ using the general relation $\Delta R = c \, \Gamma \delta \, \tau_\mathrm{fl}/(1+z)$. Next, for a given fixed $\Delta R$ we varied $R_\mathrm{start}$ and $R_\mathrm{stop}$ together with the normalization of the injection function $K_e$, evaluated the resulting $\gamma$-ray fluxes within the $0.1-300$\,GeV range for each set of the model parameters (up to the distance $R_\mathrm{end} > R_\mathrm{stop}$), and fitted the observed {\it Fermi}-LAT light curves with the simulated profiles. The final values of the free parameters were then chosen based on the $\chi^2$ of the model fits to the decaying phases of the flares. The results of the simulations are presented in Figure\,\ref{F:lcmodelf1}. 

\begin{figure*}[!t]
\begin{center}
 \includegraphics[width=\textwidth, bb=30 40 880 370,clip]{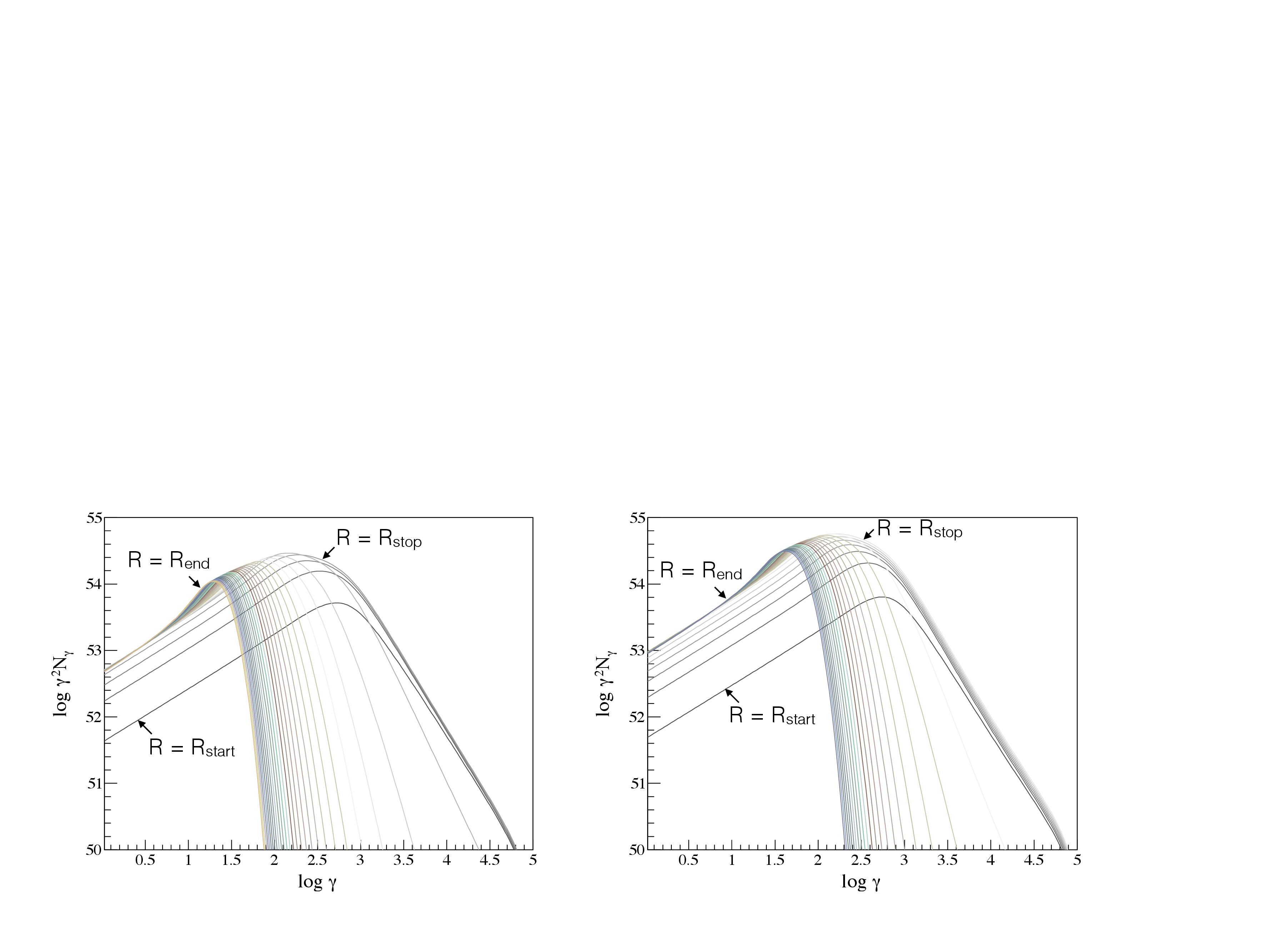}
\caption[]{The simulated evolution of the electron energy distribution as the emitting shell propagates along the free-expanding jet during the analyzed Flare\,1 and Flare\,2 (left and right panels, respectively), corresponding to the sets of the model parameters given in Table\,\ref{T:modelparam} .}
\label{F:ng}
\end{center}
\end{figure*}

In the case of the Flare\,1, the best model fit to the data returns $R_\mathrm{start} = 0.7 \times 10^{18}$\,cm and $R_\mathrm{stop} = 0.9 \times 10^{18}$\,cm, with the $\gamma$-ray emission settling down around $R_\mathrm{end}= 2.3\times 10^{18}$\,cm with the uncertainty of $0.1\times 10^{18}\,\mathrm{cm}$. Similarly, for the Flare\,2 we obtain $R_\mathrm{start} = 2.3 \times 10^{18}$\,cm, $R_\mathrm{stop} = 3.4\times 10^{18}$\,cm, and $R_\mathrm{end} = 6.9\times 10^{18}$\,cm, with the uncertainty of $0.3\times 10^{18}\,\mathrm{cm}$. Overall, the studied $\gamma$-ray flare light curves can be fitted reasonably well under all the model assumptions specified above, although the well-resolved rising profiles of the flares (especially of the Flare\,2) seem to suggest a more complex time-dependence of the electron injection function at the very beginning of the acceleration process.

The simulations performed allow us to propagate the evolution of the electron and the broad-band emission spectra during the analyzed epochs (though it should be noted at the same time that the photon statistics of the available {\it Fermi}-LAT data precludes us from any precise characterization of the $\gamma$-ray spectral changes in the source in three-hour bins). Figure\,\ref{F:ng} presents the results of the simulations regarding the evolution of the electron energy distribution as the emitting shell moves along the free-expanding outflow. As shown, until the injection stops, the number of the injected relativistic electrons grows. After the injection terminates, higher energy electrons cool very rapidly, so that the break energy in the evolved electron spectrum decreases with time, and the high-energy continuum steepens. At the later stages of the evolution ($R \sim R_\mathrm{end}$), a moderate pile-up can be noted around electron energies $\gamma \lesssim 100$. The reason for this pile-up is however not the Klein-Nishina suppression of the inverse-Compton scattering efficiency \citep[see][]{Dermer02,Moderski2005}, but instead the fact that at such late evolutionary stages the low-energy segment of the electron distribution, characterized by the injection spectral index $p < 2$, becomes subjected to the efficient radiative cooling $\mathrm{d}\gamma/\mathrm{d}t^{\prime} \propto \gamma^2$ \citep[see in this context][]{Kardashev62}. We also note that no pronounced spectral hardening due to the Klein-Nishina effects at higher electron energies is seen in our simulations, because the flare is produced effectively outside the BLR \emph{and} the injected high-energy segment of the electron continuum is steep \citep[c.f.][]{Cerruti13}; we have however confirmed that cooled electron spectrum becomes softer above electron Lorentz factors $\gamma \sim 10^3$ when Klein-Nishina supression is ignored.

\begin{figure*}[!t]
\begin{center}
\includegraphics[width=\textwidth, bb=20 10 950 410,clip]{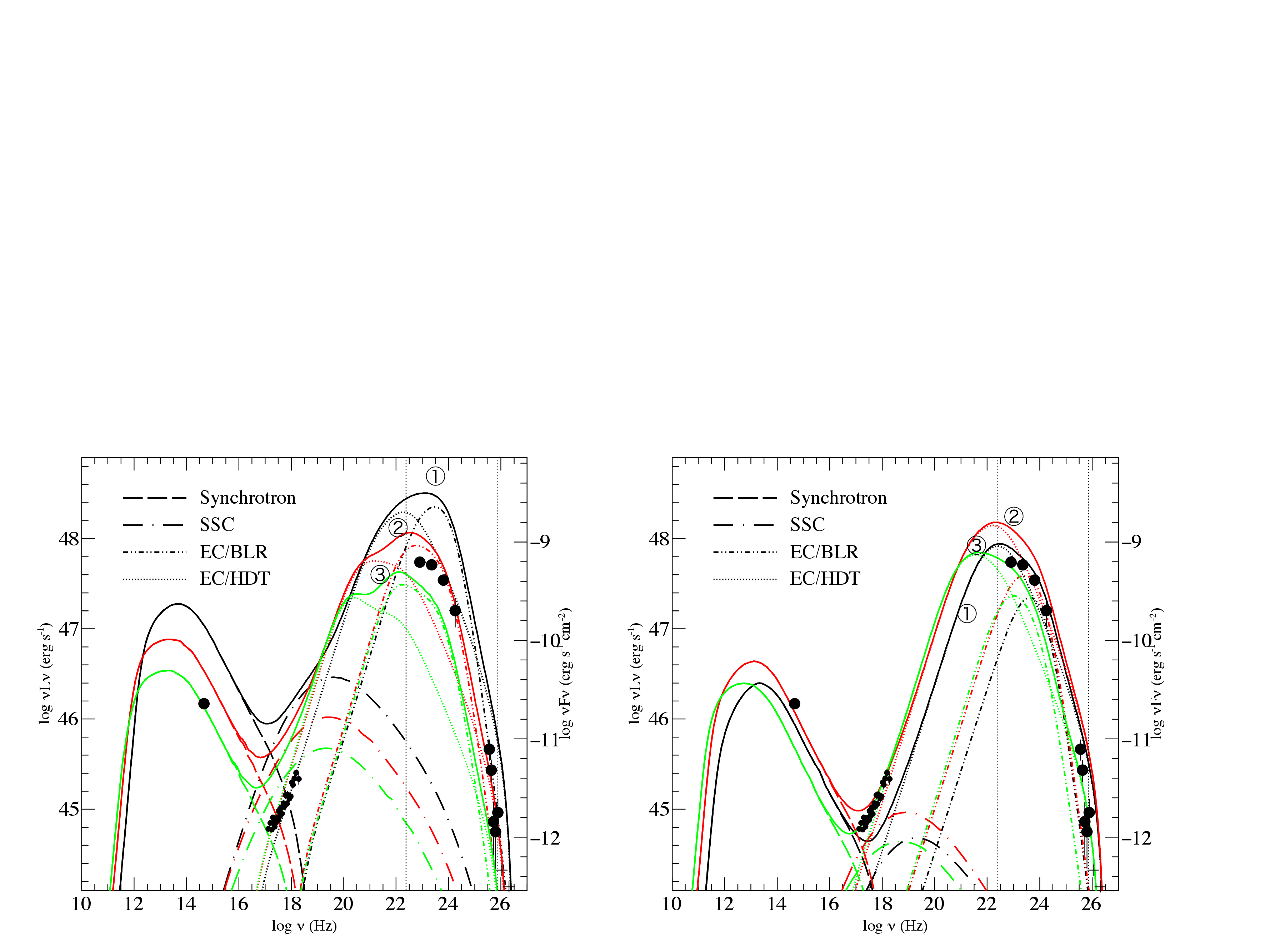}
\caption[]{The simulated time evolution of the broad-band SEDs of
 PKS\,1510--089 during the analyzed Flare\,1 and Flare\,2 (left and
 right panels, respectively). The spectra were extracted at the instants marked as ``1'', ``2'', and ``3'' in Figure\,1 (black, red, and green curves, respectively). Vertical dashed lines indicate the energy range of 100\,MeV -- 300\,GeV. For a reference, in the plots we include also the archival spectral datapoints corresponding to the March 2009 flare of the source from \citet{Barnacka2014}. Calendar dates are October 19.5--20.5, 2011 for Flare\,1 and November 7--9, 2011 for Flare\,2. }
\label{F:seds}
\end{center}
\end{figure*}

The corresponding evolution of the broad-band jet flaring spectra is shown in Figure\,\ref{F:seds}, including the synchrotron, synchrotron self-Compton (SSC), and the two ``external-Compton'' (EC/BLR and EC/HDT) emission components. As shown, in the case of the Flare\,1, the EC/HDT component dominates the production of the high-energy continuum from the observed hard X-ray up to $\sim 100$\,MeV photon energies; the soft X-ray range is dominated by the SSC emission, the GeV range by the EC/BLR process, and in the TeV range the two EC components becomes again comparable. In the case of the Flare\,2, on the other hand, the SSC and EC/BLR processes are in general much less relevant, as expected taking into account the larger size and distance of the emitting region involved in the production of this flare.

Unfortunately, no simultaneous broad-band data are available for the two flares analyzed here. \citet{Saito13} reported that the photon indices within the LAT range during the flares were approximately $\simeq 2.0$, which is roughly consistent with the spectra simulated in this work given the large uncertainty range. For a reference, in Figure\,\ref{F:seds} we included the archival spectral datapoints corresponding to the March 2009 flare of PKS\,1510--089 from \citet{Barnacka2014}.

Let us finally comment on the asymmetric profiles of the modeled flux enhancements. The standard view on this issue is that longer decay phases of blazar flares when compared with the flux rising phases reflect longer radiative cooling timescales when compared with extremely rapid, almost instantaneous particle acceleration timescales, or even with the short but finite injection intervals controlled by the macroscopic (shock dynamical) timescales (as anticipated in our work). However, radiative cooling timescales in PKS\,1510--089 jet at sub-pc/pc scales are typically expected to be much shorter than the observed decay timescales of the 2011 flares \citep[see the discussion in][]{Saito13}. 

And indeed, our simulations reveal that the dominant factor shaping the flux decay profiles is not the radiative cooling of the high-energy particles, but instead a gradient of the Doppler factor across the emitting shells of the jet plasma. Namely, for the model parameters considered above in this section, the emitting shells are relatively thin but instead considerably extended in the radial direction (see \S\,\ref{S:discussion}), so that different parts of the shells are observed at different viewing angles. As a result, emission produced within the parts characterized by the largest inclinations arrive to the observer with a significant delay when compared with the emission output of the parts located at smaller viewing angles.

\begin{figure}[!t]
\begin{center}
 \includegraphics[width=\columnwidth, bb=10 10 530 520,clip]{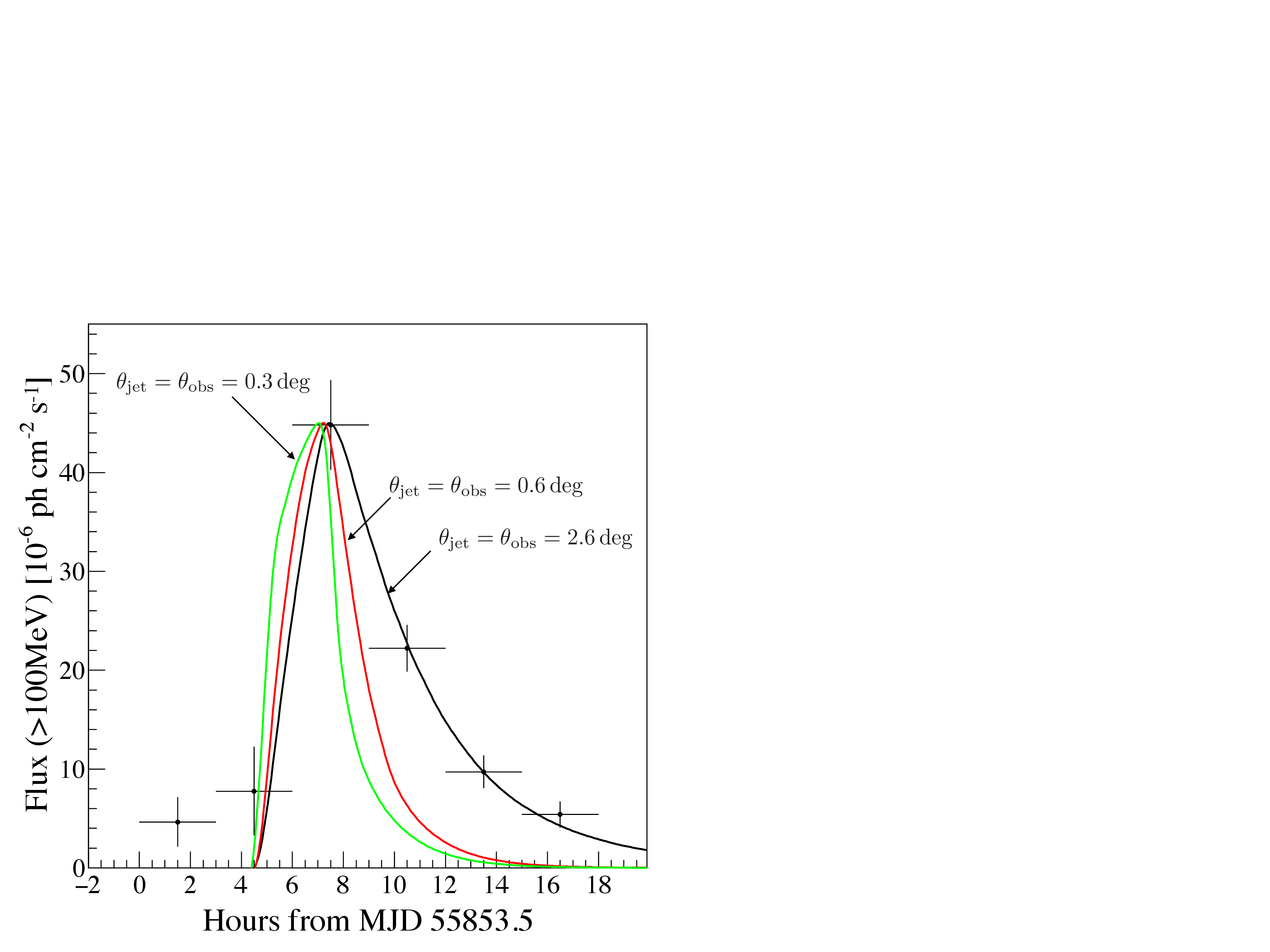}
\caption[]{Simulated profiles of the Flare\,1 assuming different values of the jet opening angle (and other parameters as given in Table\,\ref{T:modelparam} except for the adjusted electron normalization), which illustrate the effect of the Doppler factor gradient across the emitting shell in the PKS\,1510--089 jet (see \S\,\ref{S:freejet} for the discussion).}
\label{F:f1_differenttheta}
\end{center}
\end{figure}

This effect is visualized in Figure\,\ref{F:f1_differenttheta}, where we present the simulated profiles of the Flare\,1 corresponding to the different values of the jet opening angle (equal by assumption to the jet viewing angle), and all the other model parameters fixed as before except for the adjusted electron normalization. As shown, with the decreasing jet opening angle the variance of the Doppler factor across the emitting shell becomes smaller, and as a result the flare's asymmetry decreases. Only for the dramatically small $\theta_\mathrm{jet} = \theta_\mathrm{obs} = 0.3$\,deg the simulated decaying timescale becomes comparable to the radiative cooling timescale of the $\gamma$-ray emitting electrons.

\subsection{``Collimated Jet'' Model}
\label{S:collimatedjet}

Due to the causality requirement, for a free-expanding jet one has $\Gamma \, \theta_\mathrm{jet}\simeq 1$ at most, the condition which is often anticipated in blazar modeling. However, several recent radio studies of relativistic jets in blazar sources imply highly collimated outflows on milli-arcsec scales, with small opening angles $\Gamma \, \theta_\mathrm{jet} \simeq 0.1$ \citep[e.g.,][]{Clausen13,Jorstad05,Zdziarski15}. In order to investigate such a possibility in more detail, we repeated simulations of the selected flares in the source assuming different values of $0.1 \leq \Gamma \, \theta_\mathrm{jet} \leq 1$ (for the fixed $\Gamma=22$). We again imposed the condition $\theta_\mathrm{obs} = \theta_\mathrm{jet}$, so that the relativistic beaming is maximized; this requirement is necessary, since in the case of a misaligned PKS\,1510--089 jet (i.e., for $\theta_\mathrm{obs} > \theta_\mathrm{jet}$), unrealistically large electron injection is needed in order to produce the observed $\gamma$-ray flux enhancements.

The resulting best-fit positions of the onset of the $\gamma$-ray emitting regions for different values of $\theta_\mathrm{jet}$, and the corresponding normalizations of the electron injection, are presented in Figure\,\ref{F:theta_free}. For both flares considered, the location of the emission zone increases further away from the SMBH as the jet opening angle decreases. This dependance may be understood by considering the effect of the Doppler factor gradient discussed in the previous section. In particular, for smaller and smaller $\theta_\mathrm{jet}$, the Doppler factor variance across the emitting shell decreases; as a result, the emission region have to be placed further and further away from the jet base so that the increased electron cooling timescales can support the extended decay phases of the flares (note that at larger distances from the core the energy densities of the BLR and HDT photon fields decrease, and hence the corresponding radiative cooling timescales increase).

In the case of the highly collimated jet with particularly small
$\theta_\mathrm{jet} \simeq 0.26$\,deg, the best-fit position of the
emission zone turns out as $R_\mathrm{start} \simeq 4.3 \times
10^{18}$\,cm for the Flare\,1, and $R_\mathrm{start} \simeq 8.1 \times 10^{18}$\,cm for the Flare\,2. Together with the simulation results presented in the previous section, this indicates therefore a relatively narrow range allowed for the production of the observed $\gamma$-ray flares in PKS\,1510--089, namely $10^{19}$\,cm\,$\geq R_\mathrm{start} \geq 10^{18}$\,cm for the jet opening angle $0.1 \leq \Gamma \, \theta_\mathrm{jet} \leq 1$. At the same time, the required amount of the electron injection for producing the observed $\gamma$-ray luminosity may be substantially smaller for a well-collimated outflow (see the lower panels in Figure\,\ref{F:theta_free}). This means that in the case of a jet with $\Gamma \, \theta_\mathrm{jet} < 1$ it is ``easier'' to produce huge $\gamma$-ray outbursts with the total released radiative power of the order of the accretion disk luminosity, while somewhat extreme conditions are needed for such a maximum efficiency in the case of a standard, free-expanding jet with $\Gamma \, \theta_\mathrm{jet} \simeq 1$ (see the discussion in the following section).

\section{Discussion}
\label{S:discussion}

In the framework of the internal shock scenario, a collision of two plasma blobs characterized by different bulk Lorentz factors $\Gamma_1 \equiv (1-\beta_1^2)^{-1/2}$ and $\Gamma_2 \equiv (1-\beta_2^2)^{-1/2}$ produces the double-shock shock structure propagating within the merged portion of the jet matter. It is assumed that the electrons are accelerated at the shock fronts, and are injected with a given energy spectrum into the downstream region where they cool radiatively and adiabatically. The resulting shock velocity can be found under the specific assumptions on the dynamics of the colliding plasmoids. In particular, if the colliding portions of the jet flow differ only in bulk velocities, and the jet magnetic field is negligible dynamically, as assumed in this paper, the downstream (emitting) plasma is characterized by the bulk Lorentz factor $\Gamma \equiv (1-\beta^2)^{-1/2} \sim \sqrt{\Gamma_1 \, \Gamma_2}$ as long as $\Gamma_2 > \Gamma_1 \gg 1$ \citep[e.g.,][]{Stawarz04,Moderski2004}. In the upstream region rest frame (denoted below by double-primes), moving with the velocity $\beta_1$ relative to the observer, the bulk Lorentz factor of the shock is
\begin{equation}
\Gamma^{\prime\prime}_\mathrm{sh} = \sqrt{\frac{(\Gamma^{\prime\prime}+1)(4\,\Gamma^{\prime\prime}-1)^2}{8\,\Gamma^{\prime\prime}+10}} \, ,
\end{equation}
where
\begin{equation}
\beta^{\prime\prime} = \frac{\beta - \beta_1}{1 - \beta\,\beta_1} \, .
\end{equation}
This implies that the resulting shock velocity in the downstream region rest frame (denoted in this paper by primes),
\begin{equation}
\beta^{\prime}_\mathrm{sh} = \frac{\beta^{\prime\prime}_\mathrm{sh} - \beta^{\prime\prime}}{1 - \beta^{\prime\prime}_\mathrm{sh}\,\beta^{\prime\prime}} \, ,
\end{equation}
saturates at the approximately constant value of $\beta^\prime_\mathrm{sh} \sim 0.1$ \citep{Stawarz04}.

The observed rising timescale of a flare, $\tau_\mathrm{fl}$, may be
identified with the time interval $\tau_\mathrm{inj}$ when active shocks
propagating through the outflow inject freshly accelerated electrons
into the downstream region (which is a moving portion of the jet
matter). Hence, the shell width in the emitting region rest frame is
\begin{equation}
 \ell^{\prime}_\mathrm{sh} = 2\, c \beta^{\prime}_\mathrm{sh} \, \tau^{\prime}_\mathrm{inj} \simeq 2\, c \beta^{\prime}_\mathrm{sh} \, \delta \, \tau_\mathrm{fl} \, .
 \label{eq:ell}
\end{equation}
On the other hand, the transverse extension of the emitting region can be estimated simply as
\begin{equation}
r^{\prime} = r \simeq R_\mathrm{stop} \, \theta_\mathrm{jet} \, .
\end{equation}

\begin{figure}[!t]
\begin{center}
 \includegraphics[width=\columnwidth, bb=10 10 440 430,clip]{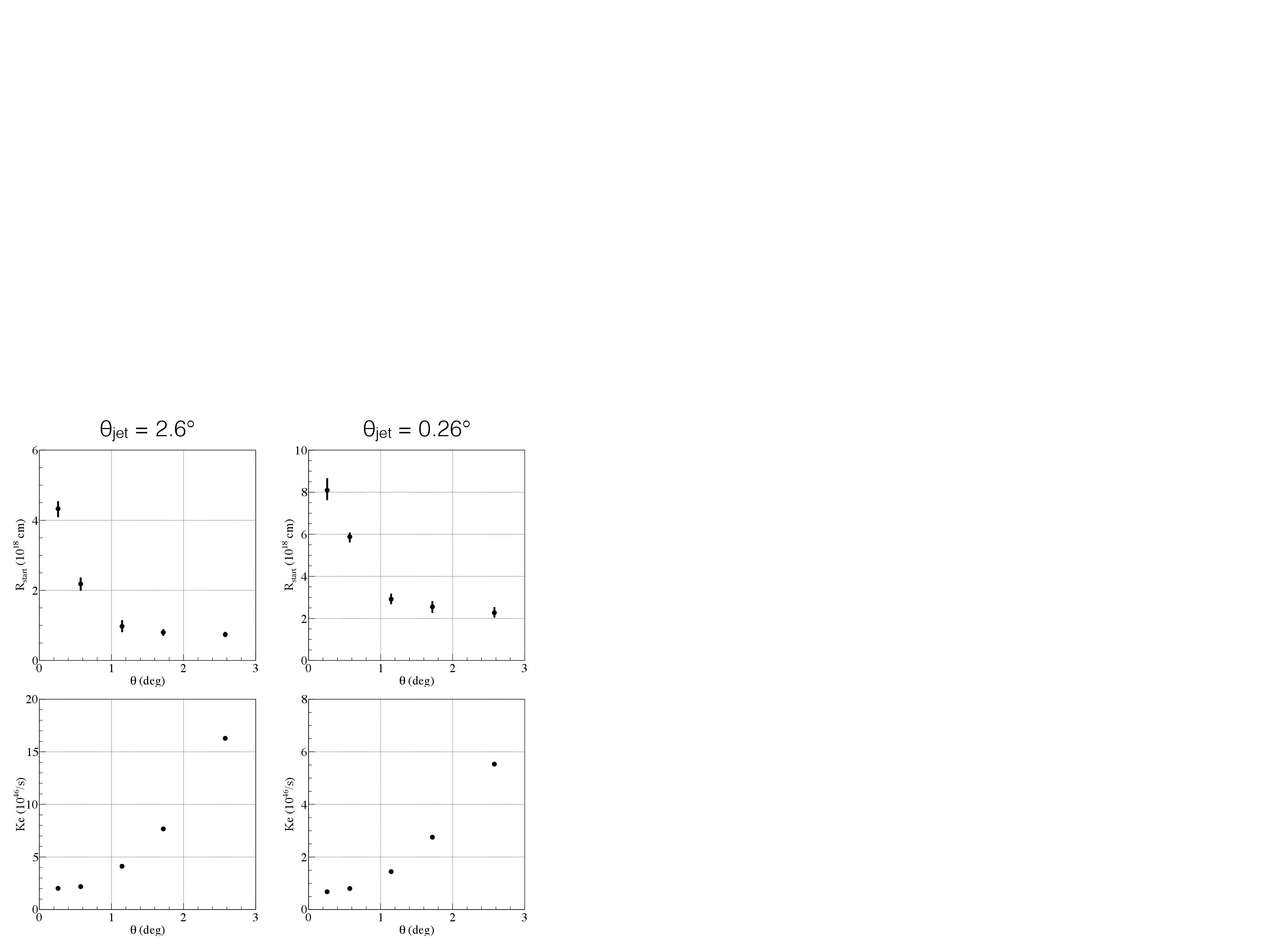}
\caption[]{The best-fit location of the onset of the flaring zone (upper panels) and the corresponding normalization of the electron injection function (lower panels) for the Flare\,1 and Flare\,2 (left and right panels, respectively), as functions of the jet opening angle $\theta_\mathrm{jet} = \theta_\mathrm{obs}$ (see \S\,\ref{S:collimatedjet} for the discussion).}
\label{F:theta_free}
\end{center}
\end{figure}

In the case of the Flare\,1, for which the observed rising timescale is of the order of three hours, one has for example $\ell^\prime_\mathrm{sh} \sim 1.5 \times 10^{15}$\,cm and $r^{\prime} \simeq 4 \times 10^{16}$\,cm under the condition of a free-expanding jet with $\theta_\mathrm{jet} = \theta_\mathrm{obs} =1/\Gamma \simeq 2.6$\,deg; the emerging ratio $\ell^\prime_\mathrm{sh}/r^{\prime} \sim 0.035$ implies therefore an extremely thin shell of the emitting plasma, even at the final stages of the shock evolution. Note, however, that in the observer frame an almost spherical apparent shape would be measured, namely $\ell_\mathrm{sh}/r = \delta \, \ell^\prime_\mathrm{sh}/r^{\prime} \sim 0.8$. The intrinsic aspect ratio of the emitting region increases, on the other hand, in the case of a well-collimated jet. In particular, with the jet opening and viewing angles $\theta_\mathrm{jet} \simeq \theta_\mathrm{obs} \simeq 0.26$\,deg, the downstream region linear scale reads as $\ell^\prime_\mathrm{sh} \sim 3 \times 10^{15}$\,cm, and the emitting region radius becomes $r^{\prime} \simeq 4 \times 10^{15}$\,cm, leading to $\ell^\prime_\mathrm{sh}/r^{\prime} \sim 0.7$, or $\ell_\mathrm{sh}/r \sim 30$ in the observer frame. In the case of the Flare\,2, characterized by the observed flux rising timescale of the order of 12 hours, the analogous values read as $\ell^\prime_\mathrm{sh} \sim 6 \times 10^{15}$\,cm and $r^{\prime} \simeq 1.5 \times 10^{17}$\,cm for $\theta_\mathrm{jet} \simeq 2.6$\,deg, or $\ell^\prime_\mathrm{sh} \sim 10^{16}$\,cm and $r^{\prime} \simeq 1.5 \times 10^{16}$\,cm for $\theta_\mathrm{jet} \simeq 0.26$\,deg, so that the corresponding aspect ratio are basically the same as for the Flare\,1.

The total number of electrons injected into the emitting shell during the shock activity period,
\begin{equation}
\mathcal{N}_e = \int\!\!\mathrm{d}t^{\prime}\!\int\!\!\mathrm{d}\gamma \, Q_{\gamma} = \tau^{\prime}_\mathrm{inj} \times \int\!\!\mathrm{d}\gamma \, Q_{\gamma} \, ,
\end{equation}
together with the electron mean energy,
\begin{equation}
\langle \gamma_\mathrm{inj} \rangle = \frac{\int\!\!\mathrm{d}\gamma \, \gamma \, Q_{\gamma}}{\int\!\!\mathrm{d}\gamma \, Q_{\gamma}} \, ,
\end{equation}
gives the jet comoving energy density of the radiating electrons
\begin{equation}
u^{\prime}_e = \frac{\langle \gamma_\mathrm{inj} \rangle \, m_e c^2 \, \mathcal{N}_e}{\pi r^2 \, \ell^{\prime}_\mathrm{sh}} \, .
\end{equation}
Hence, the electron kinetic flux (including also the power dissipated radiatively during the flare) reads as
\begin{eqnarray}
L_e & = & \pi r^2 \, c \beta \Gamma^2 \, u^{\prime}_e \\
&= & \frac{1}{2} \beta \Gamma^2 {\beta^{\prime}_\mathrm{sh}}^{-1} \, m_e c^2 \times \int\!\!\mathrm{d}\gamma \, \gamma \, Q_{\gamma} \, \nonumber
\end{eqnarray}
(see equation \ref{eq:ell}). With the best-fit parameters of the free-expanding jet model (see Table\,\ref{T:modelparam}), one obtains therefore $L_e \simeq 1.3 \times 10^{47}$\,erg\,s$^{-1}$ for the Flare\,1, and $L_e \simeq 0.5 \times 10^{47}$\,erg\,s$^{-1}$ for the Flare\,2.

With the aforementioned model parameters, the electron mean energy reads as $\langle \gamma_\mathrm{inj} \rangle \simeq 110$. In the framework of the anticipated internal shock scenario, assuming in addition that the upstream jet plasma is cold (meaning at most trans-relativistic energies of plasma particles), this value excludes a pure electron-positron content of the PKS\,1510--089 jet. That is because in the case of a pure pair plasma, the energy conservation $\langle \gamma_\mathrm{inj} \rangle \, \Gamma \, \mathcal{N}_e \, m_e c^2 \simeq \frac{1}{2} \, \Gamma_2  \, \mathcal{N}_e \, m_e c^2$ would then imply unrealistically high bulk Lorentz factor of the faster shell, namely $\Gamma_2 \sim 5000$ (for the the downstream bulk Lorentz factor $\Gamma = 22$). Hence, self-consistency of the modeling presented in this paper requires protons to dominate the plasma inertia. 

Let us therefore assume that the colliding shells with total energies $E_1$ and $E_2$ are initially dominated by cold protons with the total number $\mathcal{N}_p$. Under the $\Gamma_2 > \Gamma_1 \gg 1$ assumption specified previously, one can derive the efficiency of the energy dissipation after the shells' collision as
\begin{equation}
\eta_\mathrm{diss} = \frac{E_\mathrm{diss}}{E_1+E_2} \simeq \frac{\left( \Gamma_2/\Gamma -1\right)^2}{\left( \Gamma_2/\Gamma\right)^2 +1} \, .
\end{equation}
Quantifying next the energy dissipated per proton in the $m_p c^2$ units as
\begin{equation}
\kappa \equiv \frac{E^{\prime}_\mathrm{diss}}{\mathcal{N}_p \, m_p c^2} \, ,
\end{equation}
and keeping in mind that $E_1 = \frac{1}{2} \Gamma_1 \, \mathcal{N}_p m_p c^2$, $E_2 = \frac{1}{2} \Gamma_2 \, \mathcal{N}_p m_p c^2$, and $E^{\prime}_\mathrm{diss} = E_\mathrm{diss} / \Gamma$, one can find
\begin{equation}
\kappa \simeq \frac{1}{2} \frac{\left(\Gamma_2/\Gamma -1\right)^2}{\Gamma_2/\Gamma} \, .
\end{equation}
And since $\langle \gamma_\mathrm{inj} \rangle \, \mathcal{N}_e \, m_e c^2 = \eta_e \, E^{\prime}_\mathrm{diss}$, where $\eta_e$ is the efficiency of the energy transfer to relativistic electrons at the shock front, the composition of the jet can be finally estimated as
\begin{equation}
\frac{ \mathcal{N}_e}{ \mathcal{N}_p} \simeq \eta_e \, \kappa \,\, \frac{m_p/m_e}{\langle \gamma_\mathrm{inj} \rangle}
\end{equation}
\citep[see, e.g.,][]{Moderski2004}. The typically considered value of $\eta_e \simeq 0.5$ and a pure electron-proton jet composition $\mathcal{N}_e/\mathcal{N}_p \simeq 1$ imply therefore $\Gamma_2 \simeq 35$ with the corresponding $\eta_\mathrm{diss} \simeq 0.1$. Any larger amount of electron-positron pairs in the jet would increase both the bulk Lorentz factor of the faster shell (for the given $\Gamma$), and the overall energy dissipation efficiency; for example, $\mathcal{N}_e/\mathcal{N}_p \simeq 3$ (meaning the pair content $\mathcal{N}_{e^+}/\mathcal{N}_{e^-} \simeq 0.5$) would require a still reasonable value of $\Gamma_2 \simeq 50$, with the corresponding $\eta_\mathrm{diss} \simeq 0.25$. Note in this context that the outflows remains dominated dynamically by protons,
\begin{equation}
\frac{L_p}{L_e} = \frac{1}{\mathcal{N}_e/\mathcal{N}_p} \, \frac{m_p/m_e}{\langle \gamma_\mathrm{inj} \rangle} > 1
\end{equation}
as long as $\mathcal{N}_e/\mathcal{N}_p <10$. In particular, with $\mathcal{N}_e/\mathcal{N}_p \simeq 1-3$ adopted hereafter, one has $L_p/L_e \sim 10$.

The total luminosity of the accretion disk in PKS\,1510--089 was estimated by \citet{Nalewajko12a} as $L_\mathrm{disk} \simeq 5 \times 10^{45}$\,erg\,s$^{-1}$, meaning the accretion power in the system of the order of $L_\mathrm{acc} \sim 10^{47}$\,erg\,s$^{-1}$ (for the assumed the standard, $10\%$ radiative efficiency factor). The model parameters evaluated above under the $\Gamma \, \theta_\mathrm{jet} \simeq 1$ condition, imply therefore an extreme efficiency of the jet production, with the total jet kinetic power $L_j \simeq L_p \sim 10 \, L_\mathrm{acc}$. In the case of a highly collimated jet, this efficiency may be decreased quite significantly (see Figure\,\ref{F:theta_free}). On the other hand, the jet magnetic field within the blazar emission zone, which is rather weak already in the free-expanding jet case, becomes even less relevant dynamically with the decreasing product $\Gamma \, \theta_\mathrm{jet}$. In particular, with $\theta_\mathrm{jet} \simeq 2.6$\,deg, the ratio of the electron kinetic energy and Poynting fluxes,
\begin{equation}
\frac{L_\mathrm{e}}{L_B} = \frac{2 \, m_e c^2}{c \beta^{\prime}_\mathrm{sh}} \times \frac{\int\!\!\mathrm{d}\gamma\, \gamma \, Q_{\gamma}}{\left[R \, \theta _\mathrm{jet} \, B^{\prime}\!(R)\right]^2} \, ,
\end{equation}
reads as $L_e/L_B \sim 60$ and $\sim 20$ for the Flare\,1 and the Flare\,2, respectively. Assuming instead $\theta_\mathrm{jet} \simeq 0.26$\,deg, but keeping a comparably large Compton dominance in the source (i.e., the ratio of the high-energy and synchrotron peak luminosities $\sim 10-100$), we obtain worrisomely small jet magnetization of $L_e/L_B \gtrsim 100$. 

\begin{figure}[!t]
\begin{center}
 \includegraphics[width=\columnwidth, bb=5 5 410 770,clip]{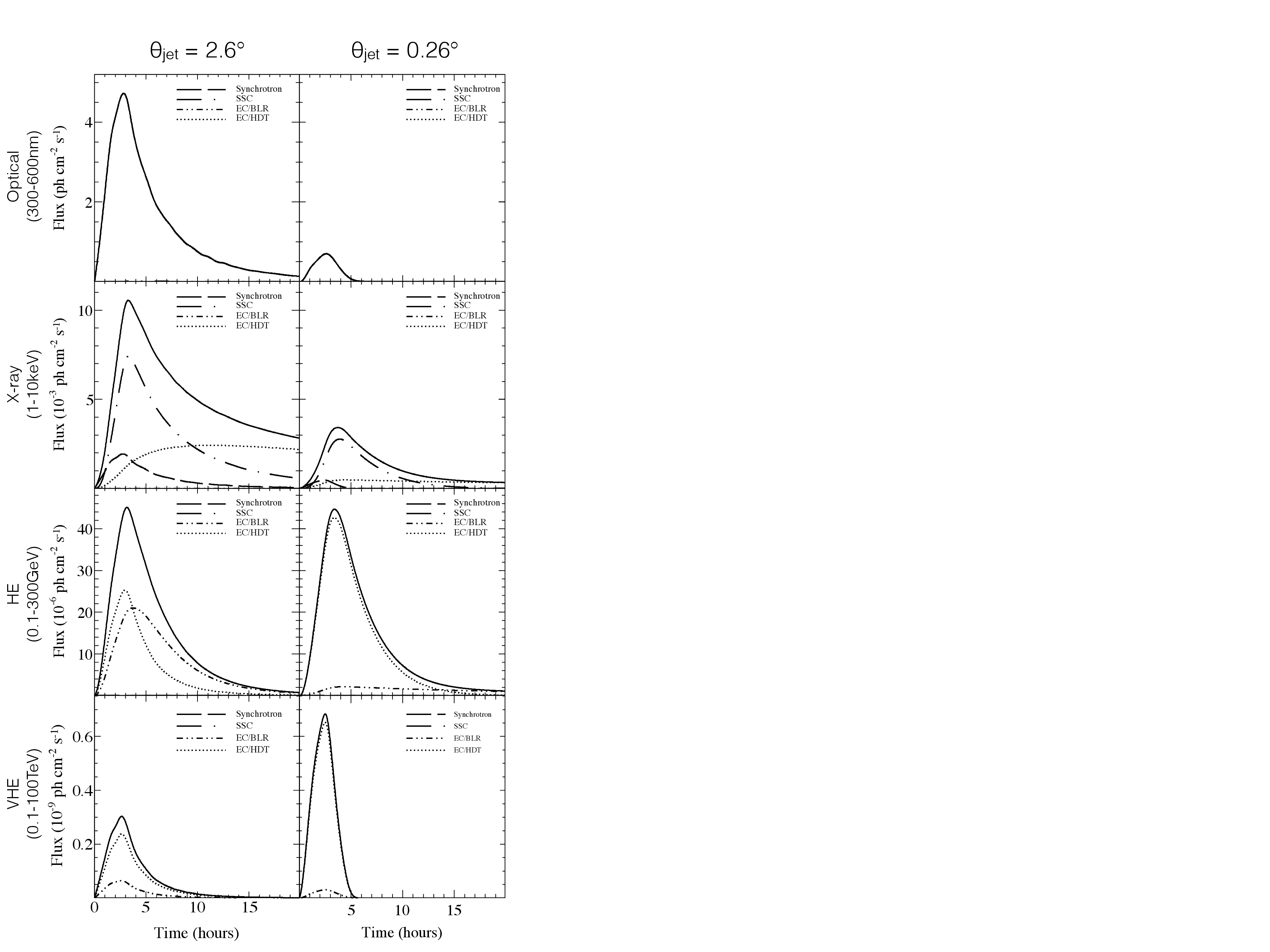}
\caption[]{Simulated light curves corresponding to the Flare\,1 in optical, X-rays, HE $\gamma$-rays, and VHE $\gamma$-rays (top to bottom, respectively), for the cases of a free expanding conical jet ($\theta_\mathrm{jet}=\theta_\mathrm{obs}=1/\Gamma$, left panels), and a collimated jet (with $\theta_\mathrm{jet}=\theta_\mathrm{obs}=0.1/\Gamma$, right panels). The location of the emission zone and the normalization of the electron injection are fixed at the best-fit values for the HE $\gamma$-ray flares (see Figure\,\ref{F:f1_differenttheta}). Various emission components are denoted by different styles of the curves (dashed, dot-dashed, dot-dot-dashed, and dotted for the synchrotron, SSC, EC/BLR, and EC/HDT, respectively), with the solid curves corresponding to the sum of all the emission components in each panel.}
\label{F:mwl_lc_f1}
\end{center}
\end{figure}

\begin{figure}[!t]
\begin{center}
 \includegraphics[width=\columnwidth, bb=5 5 410 770,clip]{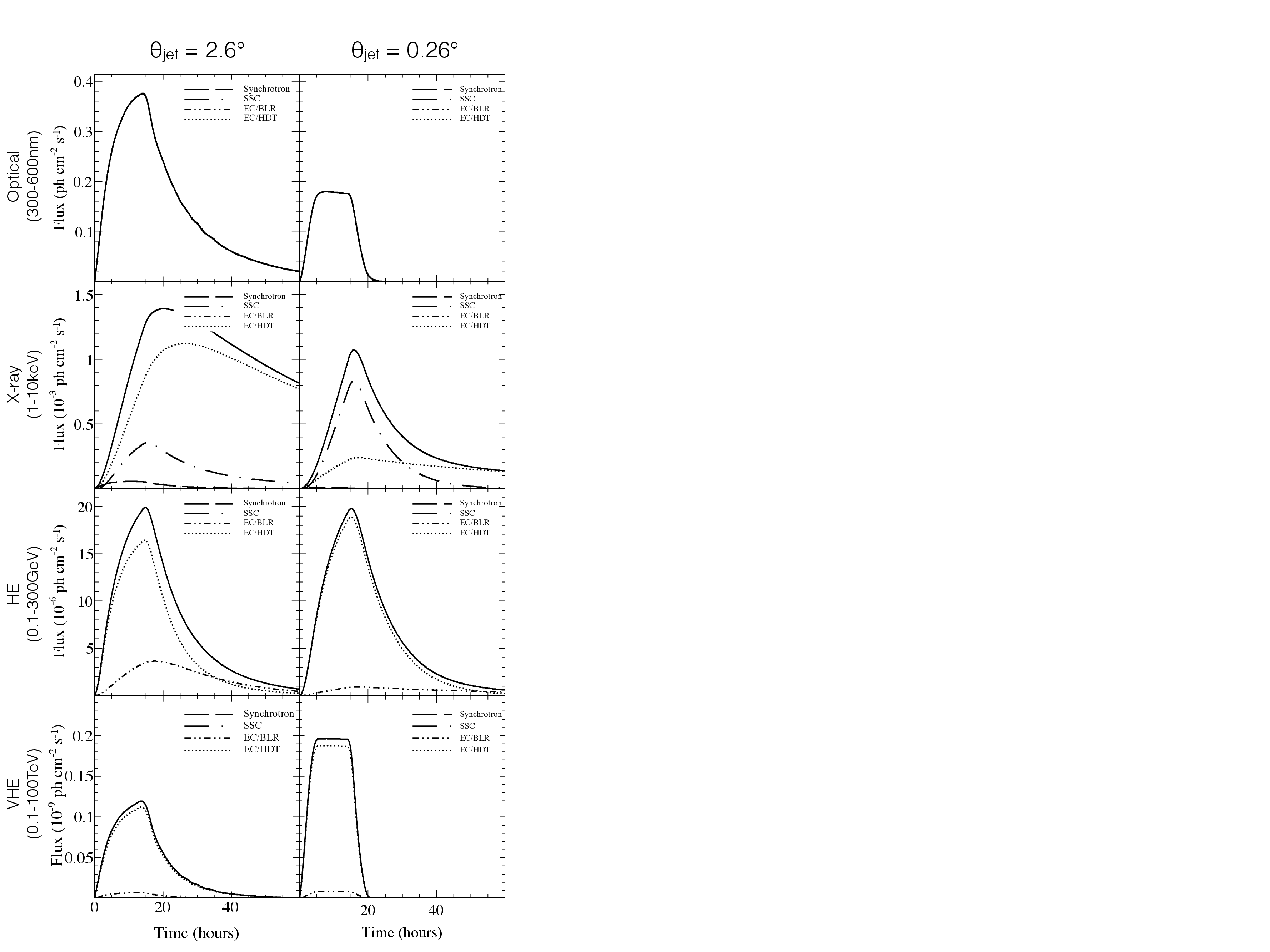}
\caption[]{Same as Figure\,\ref{F:mwl_lc_f1}, but for Flare\,2.}
\label{F:mwl_lc_f2}
\end{center}
\end{figure}

It is interesting to note in this context that the model fit to the \emph{average/quiescence} broad-band spectrum of PKS\,1510--089 using the \texttt{BLAZAR} code, presented in \citet{Kataoka08}, indicates $L_p \simeq 2 \times 10^{46}$\,erg\,s$^{-1}$ and $L_e \simeq L_B \simeq 0.6 \times 10^{46}$\,erg\,s$^{-1}$. This, together with the results of our simulations, may suggest that while during the extended source quiescence the outflow is in equipartition between the relativistic electron and magnetic field energy fluxes, with the total jet kinetic luminosity dominated only slightly by cold protons and constituting only some smaller fraction of the available accretion power ($L_\mathrm{acc} \sim 10 \, L_p$ with $L_e \sim L_B \sim 0.3 \, L_p$), dramatic though relatively rare flaring events consist of an excess energy flux carried out predominantly by the jet particles, and exceeding the mean accretion power in the source ($L_p \sim 10 \, L_\mathrm{acc}$ and $L_p \sim 10 \, L_e \gg L_B$).

The above considerations should be taken with extreme caution, however, since in the modeling presented here no synchrotron data concurrent with the analyzed $\gamma$-ray flares could be utilized. Any robust estimation of the magnetic field within the emitting region would, in fact, require exactly simultaneous infrared/optical and GeV flux measurements. At the same time, we note that small jet opening angle $\Gamma \, \theta_\mathrm{jet} <1$ implies higher number density of the radiating electrons when compared with the case of a free-expanding jet, regardless on the jet magnetization, and therefore an elevated SSC spectral component (which may dominate the entire X-ray domain, up to even the soft $\gamma$-ray regime). Hence, high-quality X-ray data simultaneous with the GeV flaring events, when modeled as presented in this paper, could, in principle, constrain robustly the jet opening angle in the source. 
 This is demonstrated more quantitatively in
Figures\,\ref{F:mwl_lc_f1} and \ref{F:mwl_lc_f2}, where we present the
simulated light curves corresponding to the Flare\,1 and Flare\,2,
respectively, in various frequency ranges, for different jet opening
angles. One can see that flaring timescales become in general shorter in
the case of a collimated outflow, the most pronouncedly however in the
X-ray domain, which constitutes a complex superposition of the SSC and
EC emission components. Also, for the given HE $\gamma$-ray flare profile and amplitude there is a significant difference between the relative amplitudes of optical/X-ray and VHE $\gamma$-ray flares depending on the jet opening angle; in particular, in the case of $\theta_\mathrm{jet}=1/\Gamma$ the optical and X-ray flares are much more prominent when compared with a moderate increase in the VHE flux, while for $\theta_\mathrm{jet}=0.1/\Gamma$ the situation reverse.

\section{Conclusions}
\label{S:conclusions}

In this paper we presented a new approach for constraining luminous blazars, incorporating fully time-dependent and self-consistent modeling of particularly bright and well-resolved $\gamma$-ray flares of PKS\,1510-089 detected with \textit{Fermi}-LAT. The two flares selected for the analysis, studied before by \citet{Saito13}, constitute the best known examples of prominent, isolated, and coherent events (with well-defined flux rising and decay phases on hourly timescales), unlike the majority of the observed blazar flux enhancements which seem rather like a superposition of distinct (though possibly related) but just unresolved sub-flaring/flickering. Unfortunately, no simultaneous data at lower (radio--to--X-ray) frequencies are available for the analyzed flares.

The results of our modeling, performed with the \texttt{BLAZAR} code developed by \citet{Moderski2003} in the framework of the internal shock scenario, are largely in agreement with the recent broad-band observations of FSRQs in general, and with the detection of TeV $\gamma$-ray photons correlated with the GeV flares in particular. Such a correlation, along with the apparent smoothness of the observed $\gamma$-ray spectra from 100\,MeV up to the TeV range, strongly suggests a co-spatiality of the GeV and TeV emitting regions, which have to be in addition located outside the BLR in order to avoid a significant attenuation of the $\gamma$-ray fluxes due to the efficient photon-photon annihilation \citep[see][]{Barnacka2014}. And indeed, the best-fit location of the $\gamma$-ray flaring region estimated here for PKS\,1510-089 turns out as $R \simeq 0.3$\,pc for a free-expanding jet with the opening angle $\theta_\mathrm{jet} \simeq 1/\Gamma \simeq 2.6$\,deg, up to $R \simeq 3$\,pc for a collimated outflow with $\theta_\mathrm{jet} \simeq 0.1/\Gamma \simeq 0.26$\,deg \citep[as advocated by][]{Clausen13,Jorstad05,Zdziarski15}. This is safely beyond the characteristic scale of the BLR in the source ($\sim 0.03$\,pc). 

We note that several other complementary arguments have been presented in the literature in support of the dominant blazar emission zone located at $\sim$\,pc distances from the core \citep[see the discussion in][]{Sikora09,Nalewajko14}.

Under the $\Gamma \, \theta_\mathrm{jet} \simeq 1$ assumption, our modeling indicates in addition an extremely efficient jet production during the flaring events and a mixed jet content, with the dominant proton energy flux exceeding the total available accretion power in the studied blazar, $L_j \sim L_p \sim 10 \, L_\mathrm{acc}$. This is in contrast to the quiescence states of the source, during which the $L_j \sim 0.1 \, L_\mathrm{acc}$ condition and an approximate equipartition between different plasma constituents (protons, electrons, and magnetic field) seem to hold. In the case of a collimated jet with $\Gamma \, \theta_\mathrm{jet} \simeq 0.1$, on the other hand, the flaring jet production efficiency decreases by an order of magnitude. Only strictly simultaneous observations of flaring PKS\,1510--089 at infrared, X-ray, and GeV photon energies, augmented by fully self-consistent and time-dependent simulations as presented in this paper, may help to remove such an order-of-magnitude uncertainty, by imposing precise constraints on the magnetization and opening angle of the emitting region.

We note that the jet production efficiency $L_j/L_\mathrm{acc}$ exceeding $100\%$ is in principle consistent with the recent understanding of the jet launching via the Blandford-Znajek process in the case when the magnetic field threading the SMBH horizon saturates at the maximum sustainable level (see \citealt{Tchekhovskoy11,McKinney12}, also the discussion in \citealt{Ghisellini14}). Yet the jet magnetization emerging in our modeling is very low, so a self-consistency of the internal shock scenario explored in this paper would require in addition a very efficient conversion of the magnetic energy flux to the particle flux (accompanied by an efficient proton loading) between the jet base and the blazar emission zone (located at $\sim$\,pc distances from the core). It may be that other energy dissipation processes are involved instead in the production of blazar flares, including turbulent acceleration \citep[e.g.,][]{Ushio10,Lefa11,Tramacere11,Cao13,Yan13,Asano14,Zheng14,Kakuwa15,Chen15}, or relativistic magnetic reconnection \citep[see, e.g.,][and references therein]{Narayan12,Giannios13,Sironi15}. These interesting alternatives have been analyzed so far more quantitatively only in the context of low-power blazars of the BL Lac type, and not FSRQs, for which the internal shock scenario is still the most appealing possibility \citep[though see in this context also the discussion in][]{Nalewajko12b}.

\acknowledgements



\L .S. was supported by Polish NSC grant DEC-2012/04/A/ST9/00083.

{}
\end{document}